\documentclass{emulateapj}

\usepackage[latin1]{inputenc}
\usepackage{graphicx}

\shorttitle{Are Ultra-long Gamma-Ray Bursts different?}
\shortauthors{Bo\"er et al.}

\begin{document}

\title{Are Ultra-long Gamma-Ray Bursts different?}

\author{M. Bo\"er, B. Gendre}
\affil{CNRS - ARTEMIS; Boulevard de l'Observatoire, CS 34229, 06304 Nice Cedex 4, France}
\email{michel.boer@unice.fr}

\and

\author{G. Stratta}
\affil{Universit\`a degli Studi di Urbino {\it Carlo Bo}, I-61029 Urbino, Italy}

\begin{abstract}

The discovery of a number of gamma-ray bursts with duration exceeding 1,000 seconds
has opened the debate on whether these bursts form a new class of sources, the so called {\em ultra-long} GRBs, or if they are rather the tail of the distribution of the standard long GRB duration. Using the long GRB sample detected by {\em Swift}, we investigate on the statistical properties of long GRBs and compare them with the  ultra-long burst properties. We compute the burst duration of long GRBs using the start epoch of the so called "steep decay" phase detected with Swift/XRT. We discuss also on the differences observed in their spectral properties. We find that ultra-long GRBs are statistically different from the standard long GRBs with typical burst duration less than 100-500 seconds, for which a Wolf Rayet star progenitor is usually invoked. Together with the presence of a thermal emission component we interpret this result as an indication that the usual long GRB progenitor scenario cannot explain the ultra-long GRB extreme duration, energetics, as well as the mass reservoir and its size that can feed the central engine for such a long time.

\end{abstract}


\section{Introduction}

Gamma-Ray Bursts (GRBs) are among the most extreme events in the Universe \citep[see the review by][]{mes06}. Their distribution of duration is very broad, as they can last for a few milliseconds to several hundreds of seconds \citep[see for instance][]{kou93}. Despite these events present an extreme diversity in terms of duration (spanning about 6 decades), variability (4 decades), energy span (8 decades when considering the afterglow phase), peak energy (2 decades), so far they have been categorized in only 2 classes \citep{dez92,kou93}. This classification is based on both their temporal and spectral statistical properties. It is this categorization that led to further studies such as the localization of the events with respect to the host galaxy \citep[e.g.][]{fon13}, pointing toward a different progenitor nature, respectively a binary system of neutron stars \citep{eic89} for the short GRBs (hereafter sGRBs) and a collapsar \citep{woo93} for the long GRBs  (hereafter lGRBs). Though finding categories in a phenomenon does not necessarily imply a different nature, this approach leads often to advances in its comprehension: a typical example is the unified model of AGNs \citep[see][for a review]{Ant} that was able to explain in a coherent picture the various manifestations of AGNs, such as Seyfert I, II, III, BL Lac, radio loud galaxies, etc. 

The collapsar model has been proposed in order to explain the amount of energy needed for a lGRB to be produced, and it has been effective in explaining several properties of these sources: e.g. the presence of a supernova \citep{hjo03, sta03} or the observation of stellar winds around the burst source \citep{gen04,gen05}. However, spectroscopic observations point toward objects with few (if any) hydrogen still present in the envelope leading to the hypothesis that the progenitor of lGRBs is a Wolf-Rayet type star \citep[e.g.][]{che99}.

Recently \citet{gen13} have proposed that GRB 111209A could not be explained by the explosion of a Wolf-Rayet star, and had properties that were markedly different from those of other GRBs, pointing possibly towards a new kind of high energy source. These so-called {\em ultra-long} GRBs (hereafter ulGRBs) last more than $10^4$ seconds. In \citet{gen13}, as well as in \citet{str13} several hypothesis for the progenitor were tested. The conclusion was that extremely massive stars, such as blue or yellow supergiant stars, can accommodate the observations. This result was later confirmed by \citet{lev13} for \object{GRB 111209A}, who added a new member to the class: \object{GRB 101225A}. More recently, Margutti et al. (2014) also showed the emergence of a new class of soft-ultra long events.

 On the other hand, \citet{vir13} who claimed that ulGRBs are rather the tail of the distribution of normal long GRBs and as a matter of consequence does not correspond to a new class of possible progenitors.

\citet{zha13} tried to estimate the actual duration of the central engine activity by modeling the overall light curve in X-rays (0.3 - 10 keV) and using a theoretical model to propose a measure of it. They define the burst duration in X-rays as the observable time over which the internal dissipation mechanisms produced inside the jet dominate the afterglow emission. They claim that the effective burst durations range continuously between 0.1 and $10^6$s instead of the usual $\rm{T}_{90}$ parameter measured in gamma-rays (15 - 150 keV for Swift) that displays a strongly peaked histogram around 30 seconds. Their conclusion is that there is no evidence for a different origin for ulGRBs, which are the tail of the distribution of lGRBs, albeit that "how to prolong a GRB central engine duration with a compact progenitor star is an open question" \citep{zha13}.

Whether ulGRBs can be accounted for by the tail of the distribution of lGRBs or not, it is rather difficult to explain a long lasting event with a compact star, and the extension of the duration of lGRBs does not solve the problem. It is therefore important to understand whether there are two players instead of one in the game. 
For example, \citet{eva14} have suggested that ulGRBs may form a distinct class of long GRBs not because of their central engine but because of their circumburst environment. By analyzing the ulGRB~130925A they formulate the hypothesis that ulGRBs reside in very low density environments that make the ejecta decelerate at times much longer than if they were in a denser medium. On the opposite, \citet{pir14} use the properties of the central engine to explain both the duration and the properties of the emission. Only one explanation is valid, and it is related to the intrinsic nature of ulGRBs.

In this article we address the question on observational grounds, splitting the problem in two parts: are the overall properties of ultra-long GRBs distinct {\it from an observational point of view} from those of normal long events, or are their properties compatible with an extension of the lGRB class toward very long durations?  In any case, whether the progenitor of ulGRB is different or not from lGRBs, a mechanism has to be found to explain them, or eventually to unify them. Should ulGRBs belong to the lGRBs class, a second question arises, that is whether the collapse of a 10-15 solar mass Wolf-Rayet type star can explain ulGRBs. We afford this question in the second part of the paper.

{We stress that the purpose of this paper is not to discuss the nature of individual bursts, but rather the nature of a class of events.

In Section \ref{sec_def}, we introduce a new measure for the burst duration in X-rays (0.3 - 10 keV), and we define ulGRBs from this measure. 
Then, in Section \ref{sec_sample}, we present our sample. In Section \ref{sec_ana} we make a statistical analysis to determine if ulGRBs can be the tail of lGRBs. We discuss the results and the need of a new class of progenitors in Section \ref{sec_discu}, before concluding. 

In the following, all errors are quoted at the one sigma level except when otherwise stated.

\section{The parameter $T_X$}
\label{sec_def}
 
The ulGRBs are
exceptional in terms of their burst duration in X-rays. More specifically, these GRBs were ultra-long in their emission duration up to the so called "steep decay phase" in the observed X-ray light curve \citep{Nou06}. The origin of this emission is likely associated to the burst "prompt emission" for which internal dissipation mechanisms were invoked, where the steep decay phase has been interpreted to mark the end of the prompt phase \citep{kum00, zha13}. 

In order to define ulGRBs we fix, somewhat arbitrary, the temporal boundary between long and ultra-long GRBs at $10^4$ seconds. However, as shown in \ref{sec_ana}, fixing a lower value, down to 1000s, does not change the conclusions, as an excess is still present.

In this work we define an empirical parameter $T_X$ as the epoch at which the steep decay phase, monitored in X-rays (0.3 - 10 keV) just after the burst trigger, starts.  This parameter can be considered as a proxy of the X-ray counterpart of the GRB burst duration typically estimated in gamma-rays (e.g. $T_{90}$). We do not consider here flares or late time steepening of the X-ray light curves, that have been suggested to be evidence of delayed central engine activity \citep[e.g.][]{zha13}, though no firm conclusions have been reached yet.

To quantify how rare are ulGRBs, we then compare the values of $T_X$ for all those long GRBs for which an estimate of this parameter was feasible from the Swift/XRT monitoring. 

We note that by using X-ray data from Swift/XRT enables to avoid two main biases that plague in general the GRB burst duration estimates and in particular the very long and ulGRBs burst duration: {\it i)} the spectral dependence of the burst duration; {\it ii)} the satellite orbit constraints, that prevent to estimate properly the duration of those GRBs longer than about 1000 s in the case of Swift. 

Indeed, the fixed X-ray energy band for all the analyzed long GRBs, secures a spectrally homogeneous  duration estimates. Moreover, Swift/XRT GRB monitoring typically starts $<100-300$s after the BAT burst trigger and lasts its first continuous observation after about 1000 s on average. Therefore, prompt emission phase analysis using Swift/XRT data is not biased against long GRBs with prompt emission lasting several hundreds of seconds. A recent analysis of the duration distribution of the time (scaled in the GRB rest frame) at which Swift's first continuous observation of each GRB ends, for which no cut-off is observed, demonstrate that observational effects do not bias significantly bursts longer than $>2000$s \citep{eva14}. 

\section{GRB sample}
\label{sec_sample}
\subsection{Building the sample}

To build the sample we use the Swift XRT GRB online catalog\footnote{http://www.swift.ac.uk/xrt\_live\_cat/} that contains Swift-XRT results for all Swift/GRBs. It provides the best fit parameters for an automated light curve analysis \citep[see details on the catalog in][]{eva09}.  To model the overall shape of the light curves multiple power law segments $f(t)=kt^{-\alpha}$ are assumed. Flare episodes are considered as extra components in the XRT GRB catalog analysis, and removed during the estimation of the power law parameters. 

\begin{figure*}
\begin{center}
\includegraphics[width=8cm]{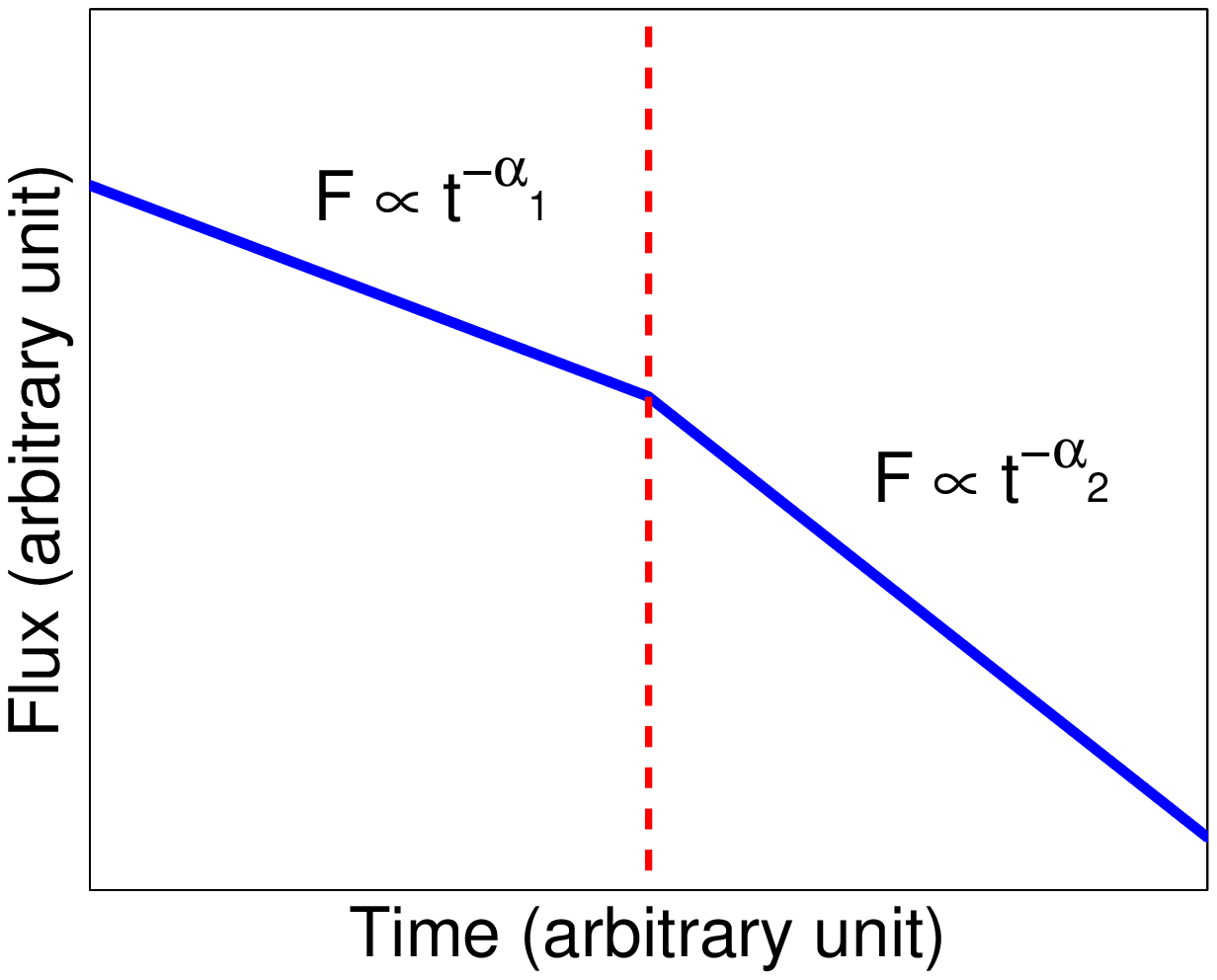} 
\includegraphics[width=8cm]{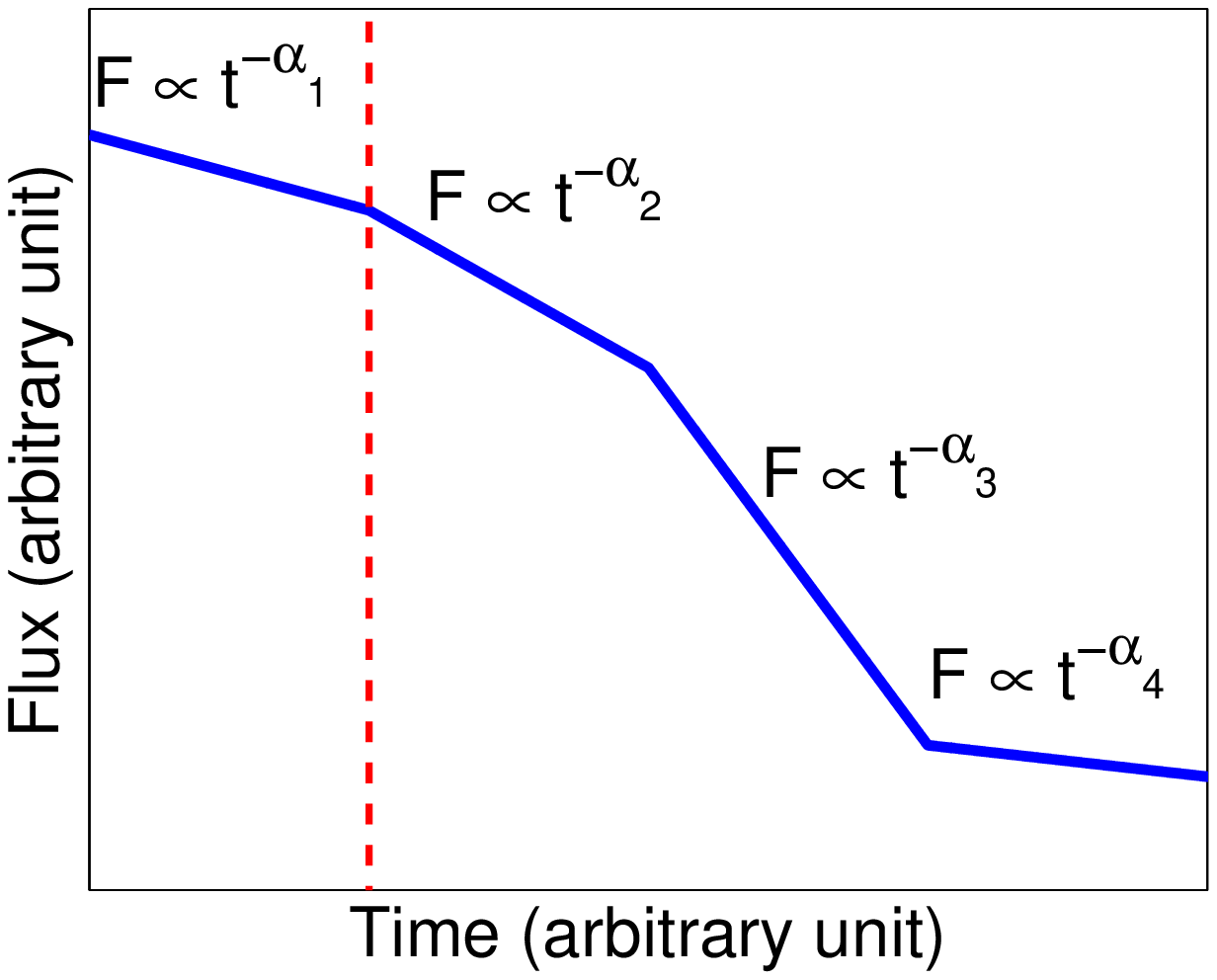}
\caption{Template light curve with the notation in use in this article. We present two example. Left: one single break. Our method allow to discriminate if this break is prompt related or jet related. Right: a complex 3-break light curve. In this example, the start of the steep decay is the second segment 
($\alpha _2 > 2.2$), 
and rule 3 (applied before rule 4) prevents to associate the start of the steep decay with the third segment. See the electronic edition of the Journal for a color version 
of this figure.
\label{fig_template}}. 
\end{center}
\end{figure*}

We prepared an automated method to extract the break time of the start of the steep decay from the above mentioned catalog. As can be seen in Fig. \ref{fig_template}, the temporal break of interest can be the first, the second, or sometime the third occurring break. We set the following rules (indicated by priority order):

\begin{enumerate}
\item a steep decay has $\alpha_i > 2.2$;
\item if $\alpha_1 > \alpha_2$ and $\alpha_1$ is steep, then the start time of the observation is larger than $T_X$;
\item if $\alpha_1 < \alpha_2$ and $\alpha_2$ is steep, then the first break time is $T_X$;
\item if $\alpha_1 < \alpha_2 < \alpha_3$ and $\alpha_3$ is steep, then the second break time is $T_X$;
\item if $\alpha_1 < \alpha_2 < \alpha_3 < \alpha_4$ and $\alpha_4$ is steep, then the third break time is $T_X$;
\item {\it et caetera}
\end{enumerate}

The application of rule 2 leads to upper limits for $T_X$. In the following we consider the start epoch of the Swift/XRT monitoring, $T_{start}$, as a $T_X$ proxy  (i.e. in case of rule 2 $T_X = T_{start}$). In order not to bias our sample we removed all bursts with a follow-up delayed by more than 500 seconds after the trigger time.

The value of the critical decay index, 2.2, is given by the electron energy distribution ($p$) of the fireball. In the standard fireball model, all segments of light curves during the afterglow part are supposed to have a decay index lower or equal to $p$. Asking $\alpha > p$ thus remove all normal cases of the fireball afterglow emission, leaving only the steep decay phase and possibly some sections post-jet break. We consider this last case by stopping at the first applicable rule, from the above ordered list (this prevents interpreting late jet breaks as the end of the prompt phase).

The final sample for which we could provide a secure $T_X$ estimate counts 207 GRBs from an original sample of 243 GRBs taken from the catalog. They are listed in table \ref{TX} (full table available in the electronic edition only).

\subsection{Peculiar events removed from the sample}

We visually inspect the BAT plus XRT light curve of our sample of GRBs. We find that on average for $T_X<500$ s the estimate of the start epoch of the steep decay phase is accurate enough for our purposes (e.g. within 5-10\%).

 For GRBs with $T_X > 500$s we found some mis-identifications of the steep decay start epoch, as explained below.

The comparison between the BAT and XRT fluxes suggest the presence of an early steep decay phase at a few 10-100 s for GRB 060813, GRB 111229A and GRB 090929B and a few 100-1000s for GRB 081029, after the trigger. For GRB 111229A, GRB 090929B and GRB 081029, this is due to a late epoch afterglow steepening not filtered out by our method. In two other cases (GRB 110422A with $T_X=1.2$ ks and 080721A with $T_X=24.5$ ks), the individuated steep decay phase was characterized by a very short duration with $\Delta T/T<0.05-0.2$ and resembling more a small flux dip in the light curve. Finally, GRB 061019 present an unusual large temporal gap between the BAT and the XRT monitoring preventing us to assess whether any earlier steep decay phase is present. We conservatively excluded all these bursts from our sample.

We now turn to the specific case of GRB 130925A \citep{eva14, pir14} which has a claim duration of more than 20ks. According to our classification this burst has a $T_X = 149$s, which does not make it remarkable.  This could be a consequence of its very strong flaring activity, that was used to claim its ultra-long origin. In addition, the BAT trigger for this burst occurred 160s after the start of the INTEGRAL observations. Therefore the prompt duration in X-rays is underestimated. Even if we take into account this delay, the resulting duration does not classify GRB 130925A as a ulGRB. Nevertheless, given the interest on this GRB and its properties, we test its addition to the sample in our analysis (see \ref{sec_ana}).

\begin{deluxetable}{cr}
\tabletypesize{\scriptsize}
\tablecaption{Our long GRB sample and the associated duration of the burst event in X-rays until the steep decay, taken from the  XRT GRB catalog best fit parameters (see text).\label{TX}}
\tablewidth{0pt}
\tablehead{
\colhead{GRB} & \colhead{$T_X$ (s)} 
}
\startdata
GRB 130609B & 453 \\
GRB 130427A & 140 \\
GRB 130315A & 160 \\
GRB 120324A &  76 \\
GRB 120213A & 166 \\
GRB 111123A & 647 \\
GRB 111121A & 111 \\
GRB 110420A &  87 \\
GRB 110119A &  82 \\
GRB 100814A & 261 \\
... & ... \\
\enddata
\tablecomments{Table \ref{TX} is published in its entirety in the 
electronic edition of the {\it Astrophysical Journal}.  A portion is 
shown here for guidance regarding its form and content.}
\end{deluxetable}



\begin{figure*}
\begin{center}
\includegraphics[width=19cm,height=8cm]{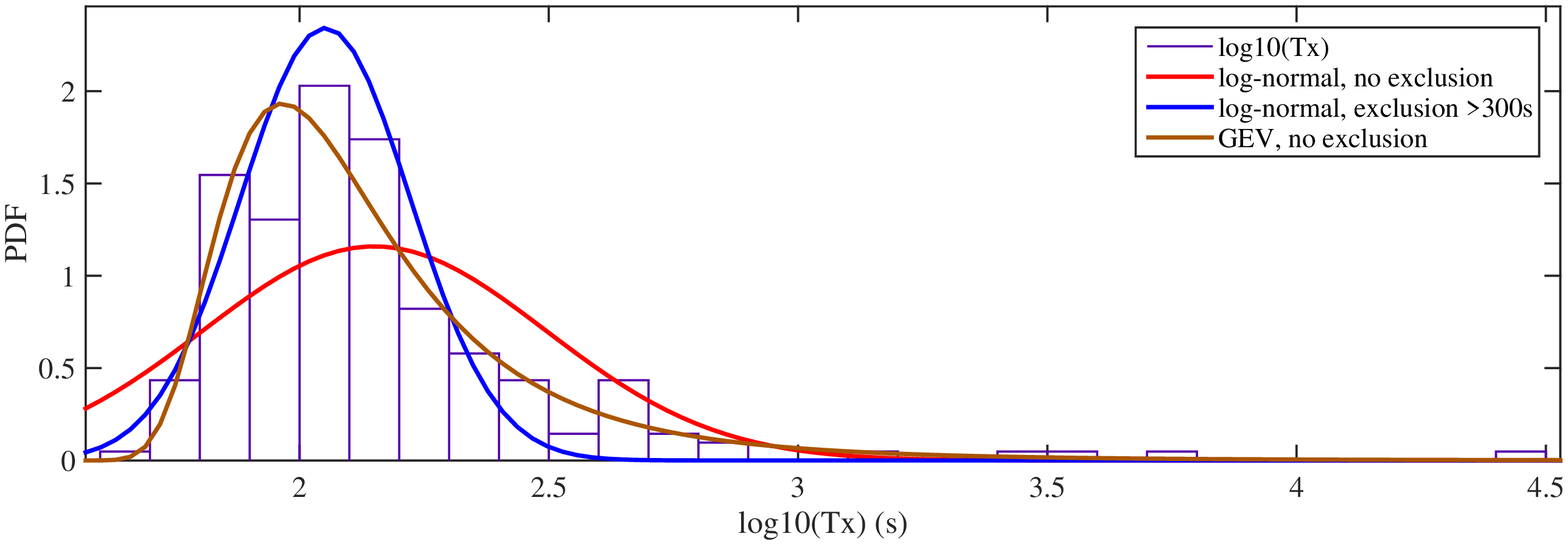} 
\caption{The probability density function of the logarithm of $T_X$. We have superimposed three different fits; red curve: log-normal model on the whole sample; blue curve: log-normal distribution, sample restricted to $T_X < 300$s; brown curve: generalized extreme value distribution, whole sample. See the electronic edition of the Journal for a color version 
of this figure.\label{fig1}}
\end{center}
\end{figure*}

\section{X-ray prompt duration distribution}
\label{sec_ana}

Our final sample is characterized by a mean $T_X$ = 337s and a median of 119 s, for a range of values that goes from a minimum of 49 s up to 25400 s. These values confirm that the X-ray burst counterpart up to the start of the steep decay phase lasts about one order of magnitude longer than the hard X-rays or gamma-ray emission duration ($\rm{T}_{90}$). This result is similar to the results reported by \citet{zha13}, although we use a different temporal parameter.  

Lets examine the question of the tail of the burst duration distribution being able to account for a few ulGRBs: for this purpose, we fit the distribution of $T_X$ with both a log-normal and a generalized extreme value (GEV) probability distribution functions, using the Matlab software package for data analysis. We use GEV because it fits better distributions with an extended tail, therefore giving a more stringent constraint on the rejection of the hypothesis that ulGRBs $T_X$ belongs to the "regular" distribution of lGRB durations.
The distribution of the logarithm of $T_X$ is plotted in Fig.\ref{fig1}, together with the fits for the log-normal distribution applied to the whole sample, to the log-normal distribution applied to the sample excluding durations larger than 300s, and the GEV applied to the whole sample. Our $T_x$ sample has 26/207 bursts (12\%) with $T_x>300$, 7 bursts (3.3\%) with $T_x>630$ s, 5 bursts (2.4\%) with $T_x>1000$ s and one (0.5\%) with $T_X>10$ ks.  

A log-normal model, applied to the whole sample, does not provide an acceptable fit of the distribution. By performing a one sample KS test we obtain a value D = $0.17$, corresponding to a null hypothesis probability of $2 \times 10^{-5}$. 
Selection effects might play a significant role in this result: for example, the small-duration left tail of the distribution is likely biased against short duration long GRBs since the Swift/XRT monitoring typically starts few 10-100 seconds after the trigger (Fig.1). However, by removing events shorter than 300s from the sample and using a truncated log-normal model (with the mean and standard deviation of the  log $T_x$ sample) we still do not obtain an acceptable fit. 

Interestingly, it is by excluding the long duration tail up to $T_x > 300$s s that we could nicely fit the distribution. For example, keeping only events for which $T_x < 300$ s, we get $\chi^2=13$ for 15 d.o.f. (while for $T_x<2000$s we got $\chi^2=150$ for 14 d.o.f.). This could indicate that our limit for ulGRB ($10^4$ s) is too conservative, and a value of $T_X > 10^3$ s should be more representative of this new class.

The GEV distribution provides a better fit than the log-normal, though non ideal because of the presence of a (small) excess of $T_X$ around 400-500s. The parameters are $\mu = 1.99 \pm 0.02$ (the location, i.e. 97.7s), the scale $\sigma = 0.19 \pm 0.01$,  and the form factor $\xi = 0.17 \pm 0.05$. Using this probability we get a prediction of 21 events above 400s, while 19 are observed. {\bf However, the same law gives }the probability to get a point above $10^4$s to be $2 \times 10^{-3}$. 

We tested the addition of GRB 130925A by adding its claimed duration of 20ks \citep{pir14} obtaining a very similar result.

These results clearly indicate the presence of an excess of detected ulGRBs with respect to the two distributions tested. Though this excess is already noticeable above 1ks, the presence of a GRB above 10ks is clearly an outlier.

\section{Discussion}
\label{sec_discu}
\subsection{The burst duration issue}

Other authors, for instance \citet{vir13}, have proposed that a duration of a few thousands of seconds was consistent with the fit of distribution of duration observed so far. This is quite understandable, as the fitting procedure is not a statistical test, but a way to approximate an actual distribution with a (given) functional. 
However, we show that the probability that they belong, as a population of several events, to the same distribution is rejected to a high level of confidence. 

We note also that \citet{vir13} have used the $t_{90}$ duration from Swift/BAT measurements (15-150 keV) while very long and ultra-long GRBs are characterized in the X-ray band (0.3 - 10 keV). In addition they computed the burst duration using either the BAT or the XRT data, making the sample inconsistent as the duration evolves strongly with the energy.

On the other hand \citet{zha13} propose a new measure of the burst duration in X-rays based on the time during which X-ray flares, taken as a proxy for the duration of the emission of the internal engine, are still emitted. Using this method they derive a distribution of burst duration that continuously spans the interval from 0.1 to $10^6$ seconds. Doing so, the duration of normal long GRBs is extended. However, this burst duration measure is based on the interpretation that internal shocks are emitted continuously during the event. 

This argument should be taken with caution. \citet{Nou06} have studied the generic X-ray light curve of the afterglow using {\it Swift} data. They find that the prompt event is rapidly followed by a steep decay, a break into a shallow decay phase, and a second break into a "normal" decay phase, with possible flares superimposed on both phases. The current interpretation of the phase after the steep decay is the start of the afterglow \citep[e.g.][]{wil06}. From that point, the central engine is not supposed to inject a significant amount of energy into the fireball, and most (if not all) of the accretion (that fuel the central engine) should have completed. Indeed it is on these considerations that we have based our burst duration definition.

It is true that late flares are sometime observed in the light curves of lGRBs. However, the interpretation of X-ray flares as witness of the central engine activity is still under debate \citep{Laz11,swe13}. Flares can be due to a renewed, or continuing activity of the central engine. They could also be due to a refreshed internal shock due to the (possible) slow velocity of the last blobs of matter ejected by the central engine. As a matter of consequence, the time of the last X-ray flare could measure the velocity of the slowest shells rather than the duration of the central engine activity. While the latter case would validate the \citet{zha13} proposed definition, the former would make it  more ambiguous as it would imply a latency time still accounted for by the measure of the burst duration, even if the central engine is not active.

We thus conclude that ulGRBs cannot be accounted for the tail of the duration distribution of the long GRBs, and that there is a {\it statistical difference} in these two populations.

\subsection{Should the progenitor be different?}

In section \ref{sec_ana} we demonstrated that lGRBs and ulGRBs are different with a large probability. Several mechanisms could explain this difference \citep{eva14, gen13}. The question is to know if the progenitors of the ulGRBs are the same than the progenitors of normal long GRBs. The duration alone will not give the answer: to make a parallel, it has been proposed for short GRBs a binary progenitor \citep{eic89} or a magnetar progenitor \citep{uso92}, both producing the same event duration. This argument has also been pointed out by \citet{zha13}, indicating that several other studies are needed before claiming for a given kind of progenitor.

The analysis we have performed here on the duration distribution showing that ulGRBs are not an extended tail of lGRBs  is not the only piece of evidence that suggest two distinct classes of events. GRB 111209A, GRB 101225A have also several specific properties that differentiate them from other lGRBs (see details in Th\"one et al. 2011 for GRB 101225A; Gendre et al. 2013, Stratta et al. 2013 for GRB 111209A; Piro et al. 2014 for GRB 130925A). For instance, the spectral properties of GRB 111209A as well as GRB 101225A present some differences with the rest of lGRBs: there is a detectable thermal emission during the prompt phase of these bursts. Normal long GRBs do not present a thermal emission, but the well known non-thermal "Band" law \citep{ban93}. Combining the unusual duration with their peculiar properties and different spectrum, we can conclude that GRB 111209A and 101225A are notably different from classical lGRBs, in line with recent results obtained by \citet{mar14}. A similar analysis has been made for sGRBs, and a GRB is considered as a member of the sGRB "class" if it is both "short" and "hard" \citep[see however ][for a detailed analysis of sGRBs in the rest frame]{Siellez2013}.

\citet{gen13}, following \citet{Woo12}, proposed a different kind of progenitor because of the difference on the duration of the central engine.

Another obvious point is then how to produce long bursts without flares that last a few tens of seconds: the usual Wolf-Rayet progenitor would then have problems to account for "short long events".

An additional indication for late time activity is the second steep decay phase which has been observed in few GRBs and whose origin remains unclear; given that there is no obvious explanation for this, we could not take these relatively rare features into account. 

In reality, one of the problem faced with the interpretation of the duration of GRB 111209A  is the fall back time of the external layers on the central black hole. In the case of ulGRBs we do see a continuous emission of the burst source for more than 6,000 s in gamma-rays, and 20,000 s in X-rays, both emissions being strongly correlated during the time when they were observed together. Though it is probably not the only possibility, accretion from a very extended source like a supergiant star is a natural hypothesis as proposed already by \citet{Woo12}

We note that an identical debate about GRB classification arose about GRB 790305b, the so-called "5th March event". When discovered, this event could be taken as compatible with the origin of other GRBs (thought at that time to originate from thermonuclear explosion on galactic neutron stars), or as the single known member of another class of event \citep{bar79, maz79}. The issue was set with the discovery, {\it 8 years later}, in 1987, of the multiple recurrences of GRB 790107, better known now as the magnetar SGR 1806-20 \citep{att87, lar87}. It is not the first time that the GRB community hesitates to recognize the specific origin of "peculiar" events: the reason of the doubts is that when applied to small samples, statistical tests cannot discriminate between a large sample of GRBs, and the few events claimed to belong to the new "class". Physics has to be applied to check whether it is possible to use the same model for the "peculiar" events or not. 

The case of ulGRBs is the same. Two events have a duration one order of magnitude longer than the longest lGRBs and their spectral properties during the prompt phase are markedly different from other "classical" long GRBs. Together, these two pieces of evidences lead us to consider these events to be at the least "peculiar", and difficult to explain within the framework of the classical Wolf-Rayet hypothesis for their origin. It is of course possible that this hypothesis still applies, but a mechanism has to be proposed to explain that extended duration. By extending the duration of all lGRBs, as proposed in \citet{zha13}, an acceptable explanation should be found for a large part of the lGRB class.

 \section{Conclusions}
\label{sec_conclu}

In this paper we showed that the properties of several bursts, such as GRB 111209A are outstandingly different from that of other lGRBs, making them representative of a new category of bursts, the ulGRBs. Does another progenitor type explain better the observations? In \citet{gen13} and \citet{str13} we proposed several possible progenitors, which all imply a larger reservoir to feed the prompt event. Any model of the origin of ulGRBs should account for a large available mass, distributed in such a way to reproduce for the extreme duration of the events. Moreover, we noted already in \citet{gen13} that the properties of these bursts make their detection very difficult, if not impossible at redshifts larger than 1. Such models should then take into account the properties of the "local" Universe, when compared to very distant events.

\acknowledgments

This paper is under the auspices of the FIGARONet collaborative network supported by the Agence Nationale de la Recherche, program ANR-14-CE33. This work made use of data supplied by the UK Swift Science Data Centre at the University of Leicester.

.

{\it Facility:} \facility{Swift}

\end{document}